\documentclass[11pt]{article}
\usepackage{latexsym}

\usepackage{amsmath}

\usepackage{amsfonts}

\usepackage{amssymb}

\newcommand{\singlespace}{\baselineskip 4.2333mm \parskip 4.2333mm}

\makeatletter
\@addtoreset{figure}{section}
\def\thefigure{\thesection.\@arabic\c@figure}
\def\fps@figure{h, t}
\@addtoreset{table}{bsection}
\def\thetable{\thesection.\@arabic\c@table}
\def\fps@table{h, t}
\@addtoreset{equation}{section}

\makeatother
\def\mathbb#1{\mathbf{ #1}}

\newtheorem{thm}{Theorem}[section]
\newtheorem{prop}[thm]{Proposition}
\newtheorem{lem}[thm]{Lemma}

\newcommand{\ms}{\medskip}

\newcommand{\noi}{\noindent}

\newcommand{\bea}{\begin{eqnarray}}
\newcommand{\eea}{\end{eqnarray}}

\newcommand{\gr}{Groenewold}
\newcommand{\vh}{Van~Hove}
\newcommand{\vn}{Von~Neumann}
\newcommand{\q}{{\cal Q}}
\newcommand{\p}{C^{\infty}(M)}

\renewcommand{\a}{{\mathcal A}}
\newcommand{\h}{{\mathcal H}}

\newcommand{\oo}{{\mathcal O}}

\newcommand{\fd}{finite-dimensional}
\newcommand{\id}{infinite-dimensional}

\newcommand{\pa}{Poisson algebra}
\newcommand{\pb}{Poisson bracket}
\newcommand{\la}{Lie algebra}
\newcommand{\lsa}{Lie subalgebra}

\renewcommand{\ss}{semisimple}
\newcommand{\sm}{symplectic manifold}

\newcommand{\ba}{basic algebra}
\newcommand{\fb}{{\mathfrak b}}
\newcommand{\fa}{{\mathfrak a}}

\newcommand{\fh}{{\mathfrak h}}
\newcommand{\fn}{{\mathfrak n}}

\newcommand{\fk}{{\mathfrak k}}
\newcommand{\fm}{{\mathfrak m}}

\newcommand{\T}{{T^*\!\,}}

\def\endproof{\hfill $\blacksquare$}

\newcommand{\C}{{\mathbb C}}
\def\r{{\mathbb R}}
\def\z{{\mathbb Z}}
\def\n{{\mathbb N}}

\begin{document}

\title{On quantizing semisimple basic algebras, I: sl$(2,\r)$}

\author{{\bf Mark J. Gotay}\thanks{Supported in part by NSF grants
DMS 96-23083 and 00-72434.}  \\
Department of Mathematics \\ University of Hawai`i \\ 2565 The
Mall \\ Honolulu, HI 96822  USA \\ {\footnotesize E-mail:
gotay@math.hawaii.edu}
}
\date{December 11, 2000 \\ (Revised November 14, 2001)}
\maketitle
\normalsize
\vspace{-0.1in}

\begin{abstract}

We show that there is a consistent polynomial quantization of the
coordinate ring of a basic nilpotent coadjoint orbit of a semisimple Lie
group. We also show, at least in the case of a nilpotent orbit in
sl$(2,\r)^*$, that any such quantization is essentially trivial.
Furthermore, we prove that there is no consistent polynomial quantization
of the coordinate ring of a basic
\ss\ orbit in sl$(2,\r)^*$.

\end{abstract}

\clearpage

\newpage





\section{Introduction}\label{intro}

We continue our study of \gr -\vh\ obstructions to quantization. Let
$M$ be a \sm, and suppose that $\fb$ is a \fd\ ``\ba '' of observables
on $M$. Given a \lsa\
$\oo$ of the \pa\ $\p$ containing
$\fb$, we are interested in determining whether $\oo$
can be ``quantized.'' (See \S\S2--3 and Gotay [2000] for the precise
definitions.) Already we know that such obstructions exist in many
circumstances: In Gotay and Grundling [1999] we proved that there are no
\fd\ quantizations of $(\oo,\fb)$ on a noncompact \sm,
for any such \lsa\ $\oo$. Based on the work of Avez [1974] or 
Ginzburg and Montgomery [2000], it is straightforward to show that there
are no quantizations of
$(C^\infty(M),\fb)$ for any compact symplectic manifold $M$ and \ba\
$\fb$. Furthermore, in Gotay, Grabowski, and Grundling [2000] we proved
that there are no quantizations of the pair
$(P(M),\fb)$ on a compact symplectic manifold, where $P(M)$ is the
\pa\ of polynomials on $M$ generated by $\fb$. 

It remains to understand the case when
$M$ is noncompact and the quantizations are \id, which is naturally
the most interesting and difficult one. Here one has little control
over either the types of \ba s that can appear (in examples they range
from nilpotent to simple), their representations, or the structure of
the polynomial algebras they generate. However, in
this context it is known from Gotay and Grabowski [2001] that there is
an obstruction to quantizing $P(M)$ when $\fb$ is nilpotent,
but that there is no universal obstruction when $\fb$ is
merely solvable.

In this paper we consider the problem of quantizing $(P(M),\fb)$ in
the other extreme case, viz. when the \ba\ is \ss.
To begin, we recall from Gotay [2000] that if a symplectic manifold $M$
admits
$\fb$ as a basic algebra, then $M$ must be a coadjoint orbit in
$\fb^*$. Unfortunately, it is difficult to determine exactly which
orbits $M \subset \fb^*$ are ``basic,'' i.e. admit $\fb$ as a \ba\
(cf. \S 2). Nonetheless, we are able to give conditions which
guarantee that various types of orbits will be basic
(Proposition~\ref{prop:char}). In particular, principal
nilpotent orbits in $\fb^*$ are basic.

We then prove in \S \ref{sec:go} that there
\emph{do} exist polynomial quantizations of certain basic orbits,
specifically the nilpotent ones:
\begin{thm}
Let $\fb$  be a \fd\ \ss\ \la, and $M$ a
basic nilpotent
coadjoint orbit in $\fb^*$. Then
there exists a polynomial quantization of $(P(M),\fb)$.
\label{thm:go}
\end{thm}

The crucial structural feature underlying
Theorem~\ref{thm:go} is that nilpotent orbits $M
\subset \fb^*$ are conical, so that
the (polynomial) ideal $I(M)$ of $M$ is homogeneous. This allows us to
split the coordinate ring of $M$ as a semidirect product
$$P(M) = (\r \oplus \fb) \ltimes P_{(2)}(M),$$
where
$P_{(2)}(M)$ is the ideal of polynomials all of whose terms are at
least quadratic. The quantization constructed in the proof of
Theorem~\ref{thm:go} has the property that it is zero on
$P_{(2)}(M)$, and so is ``essentially trivial.'' We then show that
\emph{any} polynomial quantization of a nilpotent orbit in sl$(2,\r)^*$
must be essentially trivial (Proposition~\ref{prop:at}). Thus, while
polynomial quantizations of basic nilpotent orbits do exist,
this example indicates that they are likely to be uninteresting.

If $I(M)$ is
not homogeneous, then one might expect that there is an obstruction to
quantizing $P(M)$, cf. Gotay [2000]. We show in \S3 that this is
indeed the case when $\fb = {\rm sl}(2,\r)$.
Thus polynomial quantizations are forced to be trivial for 
nilpotent orbits in ${\rm sl}(2,\r)^*$, and are genuinely obstructed for
all other basic orbits.

\ms

{\bf Acknowledgements.} \ I thank A. El Gradechi, J. Grabowski, B.
Kaneshige, and R. Sjamaar for many helpful discussions. I am especially
indebted to R. Brylinski for her input; in particular for providing a
proof of Proposition~\ref{prop:char}(\emph{iii}).


\begin{section}{Semisimple Basic Algebras}
\label{sec:back}

A key
ingredient in the quantization process is the choice of a  \emph{basic
algebra of observables} in the Poisson algebra $C^\infty(M)$. This is
a (real) \lsa\
${\fb}$ of $C^\infty(M)$ such that:
\begin{enumerate}
\item[{}] (B1) $\fb$ is finitely generated,
\ms
\item[{}] (B2) the Hamiltonian vector fields
$X_b,{b\in\fb}$, are complete,
\ms
\item[{}] (B3) $\fb$ is transitive and separating, and
\ms
\item [{}] (B4) $\fb$ is a minimal \la\ satisfying these requirements.
\end{enumerate}

\noi A subset $\fb \subset C^\infty(M)$ is ``transitive'' if
$\{X_b(m) \mid b\in\fb\}$ spans
$T_mM$ at every point. It is ``separating'' provided its elements
globally separate points of $M$. Throughout this paper we assume that
$\fb$ is \fd\ and \ss, and we routinely use the Killing form to
identify $\fb$ with $\fb^*$.

As previously noted, if the \sm\ $M$ admits $\fb$ as a \ba, then $M$
must be a coadjoint orbit of the adjoint group $B$ of $\fb$.
It is of interest to determine those orbits $M \subset \fb^*$ which
admit $\fb$ as a \ba. Unfortunately, this is not a straightforward
matter. For instance, let
$\fb = {\rm sl}(2,\r)$, so that the nonzero orbits are either open
half-cones, hyperboloids of one sheet, or components of hyperboloids
of two sheets. One can verify that the first two types of orbits are
basic for sl$(2,\r)$, but that the third type is not. (Instead, the
components of hyperboloids of two sheets are are basic for
subalgebras of triangular matrices.) Note that these orbits are all
principal (i.e. have maximal dimension) in sl$(2,\r)^*$.

The instances in which $M \subset \fb^*$ is guaranteed to be basic
are listed below.
\begin{prop}  Let $\,\fb$ be a \fd\  \ss\ \la, and $M \subset
\fb^*$ a nonzero coadjoint orbit. If either:
\begin{itemize}
\item[{}] {\rm (\emph{i})} $\fb$ is compact and $M$ is principal,
\item[{}] {\rm (\emph{ii})} $\fb$ is compact and simple, and $M$ is
arbitrary, or
\item[{}] {\rm (\emph{iii})} $M$ is nilpotent
and principal,
\end{itemize}
then $M$ admits $\,\fb$ as a \ba.
\label{prop:char}
\end{prop}

Before giving the proof, we make some remarks and recall several
important facts.  As the sl$(2,\r)$ example shows, neither (\emph{i})
nor (\emph{ii}) remain valid when
$\fb$ is noncompact. It also shows that (\emph{iii}) fails if
``nilpotent'' is replaced by ``\ss.'' It is easy to see that
(\emph{iii}) no longer holds if ``principal'' is deleted: Let $O$ be
a nilpotent half cone in sl$(2,\r)$. Then the nilpotent orbit
$O \times \{0\} \subset {\rm sl}(2,\r) \oplus \: {\rm sl}(2,\r)$
has ${\rm sl}(2,\r)$ as a \ba, not
${\rm sl}(2,\r) \oplus \: {\rm sl}(2,\r)$. Similarly
(\emph{ii}) fails if
``simple'' is deleted. Finally, regarding (\emph{iii}), observe that
if there is a nonzero nilpotent orbit in $\fb^*$, then $\fb$ is
necessarily noncompact.

\ms

Given a (noncompact) \ss\ \la\ $\fb$, recall that a ``standard triple''
is a trio $\{h,e_+,e_-\}$ of elements of $\fb$ satisfying the
commutation relations
\begin{equation*}
[h,e_\pm] = \pm 2e_\pm \;\; {\rm and} \;\;\; [e_+,e_-] = h.
\label{eq:cr}
\end{equation*}
Thus $\{h,e_+,e_-\}$ spans a subalgebra of $\fb$ isomorphic to
sl$(2,\r).$ The neutral element $h$ is \ss, while $e_\pm$ are nilpotent.
Given a nilpotent element $e \in \fb$, the Jacobsen-Morozov
theorem (Thm. 9.2.1 in Collingwood-McGovern [1993])
asserts that there exists a
standard triple $\{h,e_+,e_-\}$ in $\fb$ with nilpositive element $e_+
= e$.

\ms

\emph{Proof of Proposition}~\ref{prop:char}. Parts (\emph{i}) and
(\emph{ii}) are proven in \S4 of Gotay, Grabowski, and Grundling [2000],
so here we consider only the remaining case (\emph{iii}), the proof of
which has been kindly supplied by R. Brylinski.

Clearly conditions
(B1)--(B3) are satisfied, so we need only check the minimality
condition (B4). Suppose $\fa
\subset \fb$ is transitive on $M$, so that
\begin{equation}
\fb = \fa + \fb^{e}
\label{eq:+}
\end{equation}
for every $e \in M$, where $\fb^{e}$ denotes the centralizer of
$e$.

Fix a principal nilpotent
$e_+ \in M$. We first show that $e_+$ is contained in a
Borel subalgebra (``BSA'') of $\fb$. Let $\{h,e_+,e_-\}$ be a
standard triple in $\fb$ with nilpositive element $e_+$. {}From the
representation theory of sl$(2,\r)$ we see that the eigenvalues of
$ad_h$ are integral; we may therefore decompose
\begin{equation}
\fb = \bigoplus_{i\in \z} \fb_i
\label{eq:decomp}
\end{equation}
 where $\fb_i$ is the eigenspace of $ad_h$ corresponding to the
eigenvalue $i$. Since
$e_+$ is principal, the neutral element $h$ is generic, so its
centralizer $\fh = \fb_0$ is a Cartan subalgebra (``CSA'') of
$\fb$. Since furthermore $[\fb_i,\fb_j]
\subset \fb_{i+j}$, $\fk = \fh \oplus \fn$ is a BSA, where $\fn =
\bigoplus_{i>0} \fb_i$. Finally, as $[h,e_+] = 2e_+
\in \fb_2$, it follows that $\fk$ is the desired BSA.

{}From the
proof of Thm. 5 in Kostant [1963] we know that $\fb^{e_+} \subset \fn$,
which together with (\ref{eq:+}) implies that
$\fb = \fa + \fm$ for every $B$-conjugate $\fm$ of
$\fn$. We will prove this forces $\fa = \fb$.

Since $\fb = \fa + \fn$, we may write $h = h' + n$ where $h' \in
\fa$ and $n \in \fn$. So
$$h' = h - n$$
lies in $\fa$ and is generic
(since $h$ and $h'$ have the same characteristic polynomial). Thus the
centralizer $\fh'$ of $h'$ is also a CSA of $\fb$. A calculation
based on the decomposition (\ref{eq:decomp}) shows that
$\fh' \subset \fk$. This gives rise to the Levi decomposition
$$\fk = \fh' \oplus \fn.$$

We next claim that $\fa$ contains $\fh'$. Indeed, using $\fb = \fa +
\fn$ again, we see that each element $x' \in \fh'$ gives rise to an
element $x = x' - n_{x'}$ of $\fa$, where $n_{x'} \in \fn$. Since
$\fa$ is stable under $ad_{h'}$, it follows that both $x'$ and
$n_{x'}$ lie in $\fa$. (The reason is that ${\fh'}_{\!\C}$ is the zero
eigenspace of $ad_{h'}$ in $\fb_\C$ and $\fn_\C$ is the sum of nonzero
eigenspaces. So both $x'$ and
$n_{x'}$ lie in $\fa_\C$. As both $x'$ and
$n_{x'}$ are real they must belong to $\fa$.) In particular $\fa$
contains $\fh'$.

We can now finish the proof. We have the triangular decomposition
$$\fb = \fm \oplus \fh' \oplus \fn$$
where $\fm$ is the unique $ad_{h'}$-stable complement to $\fk$ in
$\fb$. By a result of Borel and Tits [1965], the two Borel subalgebras
$\fh' \oplus \fn$ and $\fm \oplus \fh'$ are
$B$-conjugate, whence their nilradicals $\fn$ and $\fm$ are as
well. Since $\fa$ contains $\fh'$, $\fa_\C$ is the direct sum of
${\fh'}_{\!\C}$ and some of its root spaces. Using $\fb = \fa +
\fn$, we see that $\fa_\C$ contains $\fm_\C$. Similarly, using $\fb =
\fa + \fm$, we see that $\fa_\C$ contains $\fn_\C$. Thus $\fa_\C =
\fb_\C$ and so $\fa = \fb.$
\endproof

\ms

 Let $\fb$ be a \la\ and $M$ a coadjoint orbit in $\fb^*$.
Consider the symmetric algebra
$S(\fb)$, regarded as the ring of polynomials on $\fb^*.$ The Lie
bracket on $\fb$ may be extended via the Leibniz rule to a Poisson
bracket on $S(\fb)$, so that the latter becomes a Poisson algebra. Let
$I(M)$ be the associative ideal in $S(\fb)$ consisting of all
polynomials which vanish on $M$ and set $P(M) = S(\fb)/I(M)$. Since
$M$ is an orbit $I(M)$ is also a Lie ideal, hence a Poisson ideal, so the
coordinate ring $P(M)$ of $M$ inherits the structure of a \pa\ from
$S(\fb)$. We denote the \pb\ on $P(M)$ by $\{\cdot,\cdot\}$.

Let
$P^k(M)$ denote the subspace of polynomials of degree at
most $k$. (When $I(M) \neq \{0\}$, $P(M)$ is
not freely generated as an associative algebra by the elements of
$\fb$. Consequently, the notion of ``homogeneous
polynomial'' is not necessarily well-defined, but that of ``degree''
is.)  In the cases when it does make sense, we let
$P_l(M)$ denote the subspace of homogeneous polynomials of degree
$l$, so that $P^k(M) = \bigoplus_{l = 0}^k P_l(M)$. We then also
introduce $P_{(k)}(M) = \bigoplus_{l \geq k} P_l(M).$  Notice that
when $\fb$ is \ss,
$P_1(M) = \fb$ and $P^1(M) = \r \oplus
\fb$.

\end{section}


\begin{section}{Quantization}
\label{sec:go}

Fix a \ba\ $\fb$ on $M$, and let $\oo$ be any \lsa\ of $C^\infty(M)$
containing $1$ and $\fb$. By a
\emph{quantization} of
$({\cal O},\fb)$ we mean a linear map
$\q$ from $\oo$ to the linear space Op($D$) of symmetric operators
which preserve a fixed dense domain $D$ in some separable Hilbert space
$\h$, such that for all $f,g \in \oo$,

\begin{enumerate}
\item[{}] (Q1) ${\mathcal Q}(\{f,g\}) =
i[{\q}(f),{\q}(g)]$,
\ms
\item[{}] (Q2) ${\mathcal Q}(1) = I$,
\ms
\item[{}] (Q3) if the Hamiltonian vector field $X_f$ of $f$ is
complete, then $\q(f)$ is essentially self-adjoint on $D$,
\ms
\item[{}] (Q4) $\q$ represents $\fb$ irreducibly,
\ms
\item[{}] (Q5) $D$ contains a dense set of separately analytic vectors
for some set of Lie generators of $\q(\fb),$ and
\ms
\item[{}] (Q6) $\q$ represents $\fb$ faithfully.
\label{def:q}
\end{enumerate}

We refer the reader to Gotay [2000] for an extensive discussion of
these definitions. We take Planck's reduced constant to be $1$.
Here we are interested in the case when $\oo = P(M)$.

Let $\a$ be the associative algebra over $\C$ generated by $I$ along with
$\{\q(b)\,|\, b \in \fb\}$, and let $\a ^k$ denote the subspace of
polynomials of degree at most $k$ in the $\q(b)$. We say that a
quantization
$\q$ of
$P(M)$ is
\emph{polynomial} if it is valued in $\a$. That ``polynomials
quantize to polynomials'' can be regarded as a generalized ``\vn\
rule,'' cf. Gotay [2000].

\ms

\emph{Proof of Theorem}~\ref{thm:go}. Let $M$ be a basic nilpotent
orbit. Since each nilpotent orbit is conical
(Brylinski [1998]), we may choose a set of generators for $I(M)$ which are
homogeneous. As a consequence, the gradation of $S(\fb)$ by degree
passes to the quotient $P(M)$. Thus the notion of homogeneous
polynomial \emph{does} make sense in $P(M)$. Furthermore, by virtue of
the commutation relations of
$\fb$, for each $l \geq 0$ the subspaces $P_l(M)$ are
\emph{ad}-invariant: $
\{P_1(M),P_l(M)\} \subset P_l(M).$ In view of this, $\{P_k(M),P_l(M)\}
\subset P_{k+l-1}(M)$, whence each
$P_{(l)}(M)$ is a Lie ideal. We thus have the semidirect sum decomposition
\begin{equation}
\label{eq:semidirect} P(M) = P^1(M) \ltimes P_{(2)}(M).
\end{equation}

Because of (\ref{eq:semidirect}), we can obtain a polynomial quantization
$\q$ of
\emph{all} of $P$ simply by finding an appropriate representation of
$P^1(M) = \r \oplus \fb$ and setting $\q(P_{(2)}(M)) = \{0\}$! To this
end, let $\tilde B$ be the connected, simply connected Lie group with \la\
$\fb$, and let $\Pi$ be a faithful irreducible unitary representation
of $\tilde B$ on a Hilbert space $\mathcal H.$ (For instance, we may take
$\Pi$ to be a generic irreducible component of the left regular
representation of $\tilde B$ on $L^2(\tilde B)$, cf. \S5.6 in Barut and 
R\c{a}czka [1986].) Let $D \subset
\mathcal H$ be the dense set of
analytic vectors for $\Pi$, and define $\pi = -i\,d\,\Pi
\!\restriction\! D$, cf. \S11.4 \emph{ibid.} Extend
$\pi$ to $P^1(M)$ by setting
$\pi(1) = I.$ Now take $\q = \pi \oplus 0$ (recall
(\ref{eq:semidirect})); then it is straightforward to verify that $\q$
satisfies (Q1)--(Q6) and so is the required quantization of
$(P(M),\fb)$.
\endproof

\ms

Note that the quantization constructed above is infinite-dimensional.
Indeed, there can be no \fd\ quantizations of a noncompact \ba\ (Gotay
and Grundling [1999]); this is a reflection of the fact that
\ss\ Lie groups of noncompact type have no faithful \fd\ unitary
representations. Furthermore, since
$\q(P_{(2)}(M)) =
\{0\}$, this quantization is essentially trivial. When
$\fb = {\rm sl}(2,\r)$ it turns out that
\emph{any} polynomial quantization is essentially trivial, as we show
after some preliminaries.

\ms

Henceforth take $\fb = {\rm sl}(2,\r)$ and let $M$ be an arbitrary
coadjoint orbit. 
It is convenient to complexify.  Define
$$
h = \left(\begin{array}{cc}
0 & -i \\
i & 0
\end{array}\right) \;\; {\rm \ and \ } \; \;
e_\pm = \frac{1}{2}\left[\left(\begin{array}{cc}
1 & 0 \\
0 & -1 
\end{array}\right) \pm 
\left(\begin{array}{cc}
0 & i \\
i & 0 
\end{array}\right)
\right].
$$
Then $\{h,e_+,e_-\}$ is a standard triple in $\fb_\C = {\rm sl}(2,\C).$
 Note that $h^2 + 4e_+e_-$ is the Casimir element for $\fb_\C$;
consequently
\begin{equation*}
h^2 + 4e_+e_- = c
\label{eq:cas}
\end{equation*}
is constant on $M$. 

Suppose
$\q$ were a polynomial quantization of $(P(M),\fb)$ on a dense invariant
domain
$D$ in an \id\ Hilbert space $\h$. By requiring $\q$ to be complex linear,
we can regard it as a ``quantization'' of $(P(M)_\C,\fb_\C)$.  {}From now
on, we abbreviate $P(M)_\C = P$, etc. We set
$H = \q(h)$ and $E_\pm = \q(e_\pm)$, and let $(\cdot,\cdot)$ denote the
anti-commutator. 
Finally, observe that $H^2 + 4(E_+,E_-)$ is the Casimir
element for the representation $\q$ of $\fb_\C$; since by axiom (Q4)
this representation is irreducible, 
\begin{equation}
H^2 + 4(E_+,E_-) = CI
\label{eq:qcas}
\end{equation}
for some fixed
constant $C$ (cf. Prop. 3 in Gotay and Grabowski [2001]).

We first establish the following technical result.

\begin{lem}
For any nonnegative integer $r$, the set of operators
$$S_r = \{H^jE_+^{\;l}, H^kE_-^{\; m}\,|\,j+l \leq r,\; k+m \leq r\}$$
forms a basis for $\a ^r.$
\label{lem:li}
\end{lem}

\emph{Proof.} We proceed by induction on $r$. The statement is
obviously true for $S_0 = \{I\}$.
Now assume $S_{r-1}$ is a basis for $\a ^{r-1}$.

Any element of $\a ^r$ can be written
\begin{equation*}
\sum_{k+l+m = r}\alpha_{klm}^rH^kE_+^{\; l}E_-^{\; m} + {\rm lower \
degree \ terms}.
\label{eq:li1}
\end{equation*}
Now observe that 
$$E_+E_- = (E_+,E_-) - \frac{i}{2}H.$$
Applying (\ref{eq:qcas}) we may use this relation to eliminate all factors
of $E_+E_-$ in the leading terms of the expression above, thereby
obtaining
\begin{equation}
\alpha_r H^r + \sum_{{j+l = r}\atop{l\geq 1}}\beta_{jl}^{\; +}H^jE_+^{\;
l} + \sum_{{k+m = r}\atop{m\geq 1}} \beta_{km}^{\;-}H^kE_-^{\; m} + {\rm
lower \ degree \ terms}
\label{eq:li2}
\end{equation}
for some coefficients $\alpha_r, \beta_{jl}^{\; +}, \beta_{km}^{\; -}$.
Together with the induction hypothesis, this shows that $S_r$ spans
$\a ^r$.

Now suppose there exist coefficients
$\alpha_r, \beta_{jl}^{\; +}, \beta_{km}^{\; -}$, not all zero, such that
the expression (\ref{eq:li2}) vanishes. We claim that without loss of
generality we may assume $\alpha_r \neq 0$. For suppose $\beta_{JL}^{\;
+}$ were the first nonzero coefficient in this expression. By taking the
commutator of the equation (\ref{eq:li2}) $=0$ with $E_-$
$L$-times, applying the commutation relations, and simplifying using
(\ref{eq:qcas}), we obtain a condition of the form (\ref{eq:li2}) $= 0$
where now the coefficient of $H^r$ is nonzero. Similarly, if
$\beta_{KM}^{\; -}$ were the first nonzero coefficient in (\ref{eq:li2}),
then taking the commutator with $E_+$ $M$-times would lead to the same
end.

Now repeatedly take the commutator of the equation
(\ref{eq:li2}) $=0$ with $H$. This yields further independent conditions
of the form  (\ref{eq:li2}) $=0$ but with no terms involving $H^r$. By
Gaussian elimination, we may then remove all terms on the left hand side
of (\ref{eq:li2})~$=0$ of the types
$\beta_{jl}^{\; +}H^jE_+^{\; l}$ and $\beta_{km}^{\; -}H^kE_-^{\;
m}$ with $j,k < r$. Thus we end up with
\begin{equation}
\alpha_rH^r  + A_{r-1} =0
\label{eq:li3}
\end{equation}
where $\alpha_r \neq 0$ and $A_{r-1} \in \a ^{r-1}$. 
Taking the commutator of (\ref{eq:li3}) with $H$ yields
$[A_{r-1},H] = 0.$ Applying the induction hypothesis, it follows
that $A_{r-1}$ can only depend upon $H$. Thus (\ref{eq:li3})
reduces to
\begin{equation*}
\sum_{k=0}^r\alpha_kH^k = 0.
\end{equation*}

Factor this equation over $\C$:
\begin{equation}
\alpha_r (H-\lambda_r)\cdots (H-\lambda_1) = 0.
\label{eq:sumhk2}
\end{equation}
As $\alpha_r \neq 0$, 
(\ref{eq:sumhk2}) implies that the range of $T_{r-1} =
(H-\lambda_{r-1})\cdots (H-\lambda_1)$ is contained in the
$\lambda_r$-eigenspace of $H$. By the induction hypothesis $T_{r-1} \neq
0$, so there exists $\psi \in D$ such that $\psi_{r-1} =
T_{r-1}\psi$ is a (nonzero) eigenvector of $H$. In view of the
irreducibility assumption (Q4), we conclude
from sl$(2,\r)$ theory (cf. Lang [1975]) that the set
$\{E_+^{\; l}\psi_{r-1},E_-^{\; m}\psi_{r-1}\,|\,l,m \in \n\}$ contains
an infinite number of eigenvectors of $H$, corresponding to distinct
eigenvalues $\lambda$. Each such $\lambda$ must satisfy
$\sum_{k=0}^r\alpha_k\lambda^k = 0$
which is impossible. Thus $\alpha_r = 0$ and
so $S_r$ is a linearly independent set.
\hfill $\blacktriangledown$ \rule{-2mm}{0mm}
\ms

We now
determine what $\q(h^2)$ must be.

\begin{lem}
$\q(h^2) = \alpha H^2 + \gamma I$, where $\alpha, \gamma \in \C$.
\label{lem:h2}
\end{lem}

\emph{Proof.} 
By assumption $\q(h^2)$ must be a polynomial of degree $r$, say, in $H,
E_+, E_-$, which by Lemma~\ref{lem:li} we may write in the form
(\ref{eq:li2}). Since $H$ commutes with $\q(h^2)$,
from Lemma~\ref{lem:li} we see that $\q(h^2)$ can only depend on $H$:
\begin{equation}
\q(h^2) = \sum_{k=0}^r\alpha_kH^k.
\label{eq:sumhk}
\end{equation}

Using (Q1) and (Q2) to quantize the classical identity 
\begin{equation*}
3h^2 - \frac{1}{2}\{\{h^2,e_-\},e_+\} = c
\end{equation*}
we obtain
\begin{equation}
3\q(h^2) + \frac{1}{2}[[\q(h^2),E_-],E_+]  = cI.
\label{eq:qid}
\end{equation}
Substituting (\ref{eq:sumhk}) into (\ref{eq:qid}) and simplifying yields
$$\left (3-\frac{1}{2}r(r+1)\right )\alpha_rH^r + {\rm lower \ degree \
terms} = cI.$$
{}From Lemma~\ref{lem:li} it follows that $\q(h^2)$ is at most quadratic
in $H$. Taking (\ref{eq:sumhk}) with $r = 2$, again substituting into
(\ref{eq:qid}) and simplifying, we obtain the advertised expression for
$\q(h^2)$, where
$\alpha =
\alpha_2$ is arbitrary and $\gamma$ satisfies
\begin{equation}
3\gamma = \alpha (s^2-1)+c.
\label{eq:gamma}
\end{equation}

\hfill $\blacktriangledown$ \rule{-2mm}{0mm}
\ms

Using (Q1) to quantize the identity
$$he_\pm = \pm\frac{1}{4} \{h^2,e_\pm\},$$
applying Lemma~\ref{lem:h2},
and simplifying, we obtain
\begin{equation*}
\q(he_\pm) = \alpha(H,E_\pm).
\label{eq:he}
\end{equation*}
In turn, using this to quantize the identities
\begin{equation*}
e_\pm^{\;\; 2} = \pm\frac{1}{2}\{he_\pm,e_\pm\},
\label{eq:ide2}
\end{equation*}
we find that
\begin{equation*}
\q(e_\pm^{\;\; 2}) = \alpha E_\pm^{\;\; 2}.
\label{eq:e2}
\end{equation*}
Similarly, upon quantizing
\begin{equation*}
e_+e_- = \frac{1}{2}\left(h^2 -\{he_+,e_-\}\right)
\label{eq:+-}
\end{equation*}
and using the formul\ae\ above, we get
\begin{equation*}
\q(e_+e_-) = \alpha (E_+,E_-) + \frac{\gamma}{2}I.
\label{eq:ef}
\end{equation*}

Next use these formul\ae\ to quantize
the classical identities
\begin{equation*}
2\{e_+^{\;\; 2},e_-^{\;\; 2}\} + \{he_+,he_-\} = ch
\label{eq:id1}
\end{equation*}
and
\begin{eqnarray*}
  \left\{(e_+ - e_-)^2,\{e_+^{\;\; 2}-e_-^{\;\; 2},h(e_+ + e_-)\}\right\}
& \hskip -2.5ex  +  \hskip -2.5ex &
\frac{3}{4}\left\{(e_+ + e_-)^2,\{(e_+ + e_-)^2,h(e_+ - e_-)\}\right\} \\
& \!\! = \!\! & \rule{0mm}{5mm} 8ch(e_+ - e_-).
\label{eq:id2}
\end{eqnarray*}
After tedious calculations and simplifications, we end up with
\begin{equation}
\alpha^2\left(C + 3\right)H = cH
\label{eq:contra1}
\end{equation}
and
\begin{equation}
\alpha^3\left(C+ 9\right)(H,E_+ - E_-) = \alpha c(H,E_+ - E_-),
\label{eq:contra2}
\end{equation}
respectively.

With these formul\ae\ in hand, we are now ready to prove
\begin{prop}
Let $M$ be a nilpotent  orbit in ${\rm sl}(2,\r)^*$. Then for
any polynomial quantization $\q$ of $(P(M),{\rm sl}(2,\r))$,
$\q(P_{(2)}(M)) =
\{0\}$.
\label{prop:at}
\end{prop}

\emph{Proof.} We first claim that
$\q(P_2) = \{0\}$. To see this, observe that since $M$ is nilpotent, the
constant $c = 0$. Since by (Q6) $H \neq 0$, (\ref{eq:contra1}) implies
that either $\alpha = 0$ or $C = -3$ in the given representation. But if
$\alpha = 0$, then  
from (\ref{eq:gamma}) we conclude that
$\q(h^2) = 0$ which, as we show below, leads to the desired
conclusion.

In the event that $C = -3$, we turn to (\ref{eq:contra2}). 
Since $(H,E_+-E_-) \neq 0$ by Lemma~\ref{lem:li}, we must again have
$\alpha = 0$. Thus in any eventuality
$\q(h^2) = 0$ and it follows from (Q1) that $\q(P_2) = \{0\},$ since $h^2$
is a cyclic vector for the adjoint action of ${\rm sl}(2,\C)$ on $P_2$
(i.e., every element of $P_2$ can be written as a sum of repeated
brackets of elements of ${\rm sl}(2,\C)$ with $h^2$, as the calculations
above show).

Finally, it is straightforward to check that $h^l$ is a cyclic vector
for the adjoint representation of ${\rm sl}(2,\C)$ on $P_l$. Since for
$l
\geq 2$
$$h^l = \frac{1}{2l+2}\big\{\{h^2,h^{l-2}e_+\},e_-\big\}$$
(recall that $c=0$), $\q(h^2) = 0$ together with (Q1) imply that
$\q(h^l) = 0$ for $l > 2$. Thus $\q(P_{(2)}) = \{0\}$.
\endproof

\ms

When $M \subset {\rm sl}(2,\r)^*$ is not nilpotent (in which case it
must be \ss ), it turns out that it is not even possible to
polynomially quantize
$(P(M),\fb)$; rather than finding that
$\q(P_{(2)}(M)) = \{0\}$, we get an outright inconsistency.
\begin{prop}
If $M$ is a basic \ss\ orbit in ${\rm sl}(2,\r)^*$, then
there is no polynomial quantization of $(P(M),\fb)$.
\label{thm:no-go}
\end{prop}

\emph{Proof}.
We mimic the proof of Proposition~\ref{prop:at}; the only difference
is  that $c$ is now nonzero.
As before, $H \neq 0$, so by (\ref{eq:contra1})
$$\alpha^2\left(C +3\right) = c.$$
In particular, since $c \neq 0$, $\alpha \neq 0$.
Since $(H,E_+ - E_-) \neq 0$, (\ref{eq:contra2}) then gives
$$\alpha^2\left(C +9\right) = c,$$
which is the required contradiction.
\endproof

\ms

Proposition~\ref{thm:no-go} is the noncompact analogue of the results
obtained in Gotay, Grundling, and Hurst [1996] for
$\fb = {\rm su}(2)$, in
which context every orbit is \ss. In fact, the only significant
difference between the analyses of \ss\ orbits in the sl$(2,\r)$ and
su$(2)$ cases is that the representations for the former are \id, while
those for the latter are
\fd . Since moreover the complexifications of these Lie algebras are the
same (viz. sl$(2,\C)$), the arguments leading from Lemma~\ref{lem:h2} to
Proposition~\ref{thm:no-go} don't distinguish between sl$(2,\r)$ and
su$(2)$. The same is true of the results in
\S2 \emph{ibid.}, which we may therefore immediately
carry over to the present context, yielding:
\begin{prop}
Let $M$ be a basic \ss\ orbit in ${\rm sl}(2,\r)^*$. Then $P^1(M) = \r
\oplus {\rm sl}(2,\r)$ is the largest \lsa\ of the coordinate ring $P(M)$
that can be consistently polynomially quan\-tized.
\label{thm:sub}
\end{prop}
Thus the obstruction to quantizing polynomial algebras on \ss\ orbits in
sl$(2,\r)^*$ is very severe: the \emph{best} one can do is quantize the
Lie subalgebra of affine polynomials!

\ms

We end this section with a discussion of the assumption that $\q$ be
polynomial. In general, when the basic algebra $\fb$ is compact (or,
equivalently, when the coadjoint orbit $M$ is compact) every
quantization of $(P(M),\fb)$ is polynomial. For then the Hilbert space
$\h$ must be finite-dimensional, and the claim follows from a
well known property of enveloping algebras, cf. Prop. 2.6.5 in
Dixmier~[1977]. Furthermore, when $\fb$ is nilpotent, it was proven that
$\q$ must be polynomial in Gotay and Grabowski [2001]. These results are
direct consequences of the irreducibility condition (Q4). However,
the analogous statement does not seem to hold for noncompact \ss\ \ba s.

To see this, we provide an
alternate version of Lemma~\ref{lem:h2}, which does not assume that $\q$
is polynomial \emph{ab initio}. For what follows, we need to be more
specific about the domain $D$. As a consequence of (Q5), 
$\q\!\restriction \fb$ integrates to a unique unitary representation
$\Pi$ of $\tilde{B}$ on
$\h$ (Cor. 1 of Flato and Simon [1973]). Naturally associated with $\Pi$
is the derived representation of $\fb$ on the domain
$C^\omega(\Pi)$ consisting of analytic vectors of $\Pi$.
We shall henceforth assume that $D \supset C^\omega(\Pi)$. Furthermore,
for the sake of simplicity, we suppose that the representation $\Pi$
drops to SL$(2,\r)$ from its double cover $\tilde B$.

Then from sl$(2,\r)$ theory (cf. Lang
[1975]) we know that (\emph{i}) the spectrum
$\Delta$ of $H$ consists of certain imaginary integers, (\emph{ii}) in
view of (Q4), for each $-in \in \Delta$ the corresponding eigenspaces
$\h_n$ are 1-dimensional, and (\emph{iii}) each eigenvector of $H$ is an
analytic vector, so that $\h_n \subset D$. Furthermore, the quantizations
of
$\fb$ are labeled by certain complex numbers $s$, and that for each
$-in \in \Delta$, there is a vector $\psi_n \in\h_n$ such that
\begin{equation}
H\psi_n =  -in\psi_n \; \; \; {\rm and} \; \; \;
E_\pm \psi_n  =  -\frac{i}{2}(s+1 \pm n)\psi_{n \pm 2}. \label{eq:x}
\end{equation}

By (Q1), both $H$ and $\q(h^2)$ commute. {}From 
observations (\emph{ii}) and (\emph{iii}) above, and the fact that
$\bigoplus_{n \in i\Delta}\h_n$ is dense in $\h$, it follows that
\begin{equation}
\q(h^2) = \xi(H)
\label{eq:xi}
\end{equation}
for some Borel function $\xi$ on the spectrum of $H$. We now
compute $\xi$.

Apply the relation (\ref{eq:qid}) to $\psi_n$; from (\ref{eq:x}) and
(\ref{eq:xi}) we get the recursion relation
\begin{eqnarray}
&  &  \hskip -50pt 3\xi_n - \frac{1}{8}\big[\left(s+(1+n) \right)
\left(s-(1+n)\right)(\xi_n - \xi_{n+2}) \;\;\;\;\; \nonumber \\
& & \;\; -
\left(s+(1-n)\right)\left(s-(1-n)\right)(\xi_{n-2} -
\xi_n)\big]   = c,
\label{eq:rr}
\end{eqnarray}
where $\xi_n$ is defined via $\xi(H)\psi_n = \xi_n\psi_n.$ It is
straightforward to check that any \emph{polynomial} solution of this
recursion relation
 is of the form
$\xi_n =  \gamma - \alpha n^2$ from which, in view
of (\ref{eq:xi}) and (\ref{eq:x}), we recover the formula derived
previously for
$\q(h^2).$ But there are other solutions of (\ref{eq:rr}) which are
transcendental: for instance, consider the discrete series representation
with $s \geq 1$ an even integer. 
\linebreak[4] 
Then $\Delta =
-i\{s+1,s+3,\ldots\}$, and with some effort one can show that the general
solution of (\ref{eq:rr}) is
$$\xi_n = \gamma - \alpha n^2 + \beta\left( (s^2 - 3 n^2 -1)
\Big[\digamma \Big(\frac{1 + n - s}{2}\Big) - 
         \digamma\Big(\frac{1 + n + s}{2}\Big) \Big]- 6ns\right),$$
where $\alpha, \beta$ are arbitrary, $\gamma$ is given by
(\ref{eq:gamma}), and the digamma function $\digamma$ is the logarithmic
derivative of the gamma function. Similar formul\ae\ hold for other
allowable values of $s$.

Thus in the case of sl$(2,\r)$ irreducibility enables one to determine
$\q(h^2)$ and then, following the template set forth after the proof of
Lemma~\ref{lem:h2}, all of $\q(P^2(M))$, and so on. But
unlike for su(2), irreducibility alone apparently does \emph{not} suffice
to guarantee that
$\q$ is polynomial. While  Proposition~\ref{thm:no-go} shows that
polynomial quantizations of
$(P(M),{\rm sl}(2,\r))$ for \ss\ $M$ cannot exist, it is unclear whether
such transcendental quantizations are similarly obstructed.

\end{section}


\begin{section}{Discussion}
\label{sec:pf}

The quantization of $(P(M),\fb)$ for $M \subset \fb$ nilpotent given above
is not the first know example of a consistent quantization: In Gotay
[1995] a full quantization of $(C^\infty(T^2)),\mathfrak t)$ was
exhibited, where $\mathfrak t$ is the \ba\ of trigonometric polynomials
of mean zero; and in Gotay and Grabowski [2001] a polynomial quantization
of $P(\T\r_+)$, with the \ba\ being the affine algebra a(1), was
constructed. This last example ``works'' for exactly the same reason the
nilpotent one does, viz. the ideal
$I(M)$ is homogeneous. However, in contrast to the case of sl$(2,\r)$ (cf.
Proposition~\ref{prop:at}), a polynomial quantization of $P(\T \r_+)$ with
\ba\ a(1) need \emph{not} be zero on $P_{(2)}$.

In fact,
a moment's reflection shows that there will exist a polynomial
quantization of
$(P(M),\fb)$ for any \ba\ $\fb$ whenever $I(M)$ is homogeneous, for then
one has the crucial splitting (\ref{eq:semidirect}).
But this construction will fail whenever
$I(M)$ is inhomogeneous so that
$P_{(2)}(M)$ is not well-defined.
It is tempting to conjecture that an obstruction to quantization exists
whenever $I(M)$ is inhomogeneous; this is borne out explicitly here in the
case of 
\ss\ orbits in 
sl$(2,\r)$ by
Proposition~\ref{thm:no-go}. This correlation is also known to hold in
all other examples that have been investigated thus far (Gotay [2000]).

The next step is to extend
Propositions~\ref{prop:at} and \ref{thm:no-go} to higher rank \ss\ \ba
s. Clearly, this necessitates using more Poisson theoretic techniques,
as opposed to the computational approach taken here. These issues are
addressed in Gotay [2001].

\end{section}


\newpage
\section*{References}

\begin{description}
\singlespace

\item Avez, A. [1974] Repr\'esentation de l'algebre de Lie des
symplectomorphismes par des op\'erateurs born\'es. \emph{C.R. Acad.
Sc. Paris S\'er. A} {\bf 279} 785--787.
 
\item Barut, A.O. and R\c{a}czka, R. [1986] {\sl
Theory of Group Representations and Applications.} Second
Ed. (World Scientific, Singapore).

\item Borel, A. and Tits, J. [1965] Groupes r\'eductifs. {\it Publ. Math.
I.H.E.S.} {\bf 27} 55--151.

\item Brylinski, R. [1998] Geometric quantization of
nilpotent orbits. {\it J. Diff. Geom. Appl.} {\bf 9} 5--58.

\item Collingwood, D.H. and McGovern, W.M. [1993] {\sl Nilpotent
Orbits in Semisimple Lie Algebras.}  (Van Nostrand Reinhold,
New York).

\item Dixmier, J. [1977] {\sl Enveloping Algebras.} (North Holland,
Amsterdam).

\item Flato, M. and Simon, J. [1973] Separate and joint
analyticity in Lie groups representations. {\it J. Funct. Anal}. {\bf
13} 268--276.

\item Ginzburg, V.L. and Montgomery, R. [2000] Geometric quantization and
no-go theorems. In: {\sl Poisson Geometry}, J. Grabowski and P.
Urba\'nski, Eds. Banach Center Publ. {\bf 51} (Inst. Mat. PAN, Warszawa),
69--77.

\item Gotay, M.J. [1995] On a full quantization of
the torus. In: 		 {\sl Quantization, Coherent States and Complex
Structures}, Antoine, J.-P. {\it et al.}, Eds.  (Plenum, New York),
55--62.

\item Gotay, M.J. [2000] Obstructions to quantization.
In: 		 {\sl Mechanics: {}From Theory to Computation (Essays in Honor
of Juan-Carlos Simo)}, J. Nonlinear Sci. Eds.  (Springer, New York),
171--216.

\item Gotay, M.J. [2001] On quantizing semisimple basic algebras, II:
The general case. Preprint.

\item Gotay, M.J. and Grabowski, J. [2001] On quantizing
nilpotent and solvable \ba s. {\it Canadian Math. Bull.}
{\bf 44} 140--149.

\item Gotay, M.J., Grabowski, J., and Grundling, H.B.
[2000] An obstruction to quantizing compact symplectic manifolds. {\it
Proc. Amer. Math. Soc.} {\bf 28} 237--243.

\item Gotay, M.J. and Grundling, H. [1999] Nonexistence
of finite-dimensional quantizations of a noncompact symplectic
manifold. In: {\sl Differential Geometry and Applications}, I.
Kol\'a\v r \emph{et al}., Eds. (Masaryk Univ., Brno), 593--596.

\item Gotay, M.J., Grundling,  H.B., and Hurst, C.A. [1996] A
\gr -\vh\ theorem for $S^2$.
{\it Trans. Amer. Math. Soc.} {\bf 348} 1579--1597.

\item Kostant, B. [1963] Lie group representations on
polynomial rings. {\it Amer. J. Math.} {\bf 85} 327--404.

\item Lang, S. [1975] SL$_2(\r)$. (Addison-Wesley, Reading).

\end{description}


\end{document}